\documentclass[aip, reprint, twocolumn, amsmath,amsfonts,amssymb]{revtex4-1}
\usepackage[colorlinks=true,linkcolor=blue]{hyperref}
\usepackage{graphicx}
\usepackage{amsmath,amsfonts,amssymb}
\usepackage{color}
\usepackage{bm}
\usepackage[mathscr]{eucal}
\usepackage{picture}
\usepackage{float}
\usepackage{subcaption}

\usepackage{graphicx}
\usepackage{amsmath,amsfonts,amssymb}
\usepackage{color}
\usepackage{bm}
\usepackage{threeparttable}
\usepackage[mathscr]{eucal}
\usepackage{picture}

\definecolor{grey}{rgb}{0.7,0.7,0.7}

\newcommand{\br}{\nonumber\\}
\usepackage{fancyvrb}
\usepackage{ulem}

\usepackage[T1]{fontenc}

\definecolor{brown}{RGB}{111,16,50}

\definecolor{purple}{rgb}{0.8,0.0,0.8}
\definecolor{pink}{rgb}{1.,0.5,0.5}

\newcommand{\ctext}[1]{\raise0.2ex\hbox{\textcircled{\scriptsize{#1}}}}

\begin{document}

\title{Analytical Nuclear Gradients for State-Averaged Configuration Interaction Singles Variants: Application to Conical Intersections}
\author{Takashi Tsuchimochi}
\email{tsuchimochi@gmail.com}
\affiliation{College of Engineering, Shibaura Institute of Technology, 3-7-5 Toyosu, Koto-ku, Tokyo 135-8548 Japan}
\affiliation{Institute for Molecular Science, 38 Nishigonaka, Myodaiji, Okazaki 444-8585 Japan}

\begin{abstract}
We derive analytical nuclear gradients for state-averaged orbital-optimized configuration interaction singles (SACIS) and its spin-projected extension (SAECIS), enabling efficient geometry optimization and minimum-energy conical intersection (MECX) searches within a low-cost CIS-based framework. The formulation employs a Lagrangian approach and explicitly removes null-space contributions in the coupled perturbed equations to ensure numerically stable gradients. For twisted-pyramidalized ethylene, both SACIS and SAECIS qualitatively reproduce the correct conical intersection topology, in sharp contrast to conventional CIS and ECIS. Benchmark calculations on twelve MECXs demonstrate that both methods reproduce geometries with mean RMSDs below 0.1~{\AA} relative to high-level reference methods.
SACIS captures the essential degeneracy through variational orbital relaxation, which alleviates ground-state Hartree--Fock (HF) orbital bias and effectively incorporates static correlation through localization effects; notably, spin projection is found to be non-essential for the qualitative description of these intersections. Overall, SACIS and SAECIS provide qualitatively reliable CX descriptions at mean-field computational cost in a black-box manner. Given their comparable accuracy and the additional overhead associated with spin projection, SACIS offers a more favorable cost-performance balance for general applications, whereas SAECIS may become advantageous when higher excited states with significant double-excitation character are involved.
\end{abstract}
\maketitle
\section{Introduction}
Conical intersections (CXs) are fundamental features of molecular potential energy surfaces, providing efficient pathways for radiationless transitions between electronic states.\cite{Yarkony96, Worth04}
They govern ultrafast internal conversion, photoisomerization, and nonadiabatic reaction dynamics in a wide range of photochemical and photophysical processes.\cite{Polli2010,Levine07,Robb00,Matsika07}
Because CXs arise from degeneracies between electronic states as functions of nuclear coordinates, their reliable theoretical description requires a balanced treatment of multiple states on an equal footing.
Multireference wave function methods, most notably state-averaged complete active space self-consistent field (SA-CASSCF), provide a rigorous framework for describing CXs, through variational orbital optimization and explicit state averaging.\cite{Roos80, SACASSCF1, SACASSCF2}
However, their steep computational cost and sensitivity to active-space selection significantly limit their applicability to larger systems. These limitations have motivated the development of computationally less demanding alternatives.\cite{RASSCF,RASSCF_conical_intersection,Slavicek10,SAREKS}

Configuration interaction singles (CIS) offers an appealing mean-field-level approach to excited states due to its conceptual simplicity and low computational cost.
Nevertheless, CIS is intrinsically incapable of describing CXs because it lacks the static correlation required to treat near-degenerate electronic states consistently.\cite{Dreuw05, Szabo1989, Failure_in_ConicalIntersection1}
This deficiency is shared by linear-response time-dependent density functional theory (TDDFT)\cite{TDDFT,Casida} and its Tamm--Dancoff approximation (TDA),\cite{Hirata99} whose failure near CXs has been extensively documented.\cite{Failure_in_ConicalIntersection1,Failure_in_ConicalIntersection2,Failure_in_ConicalIntersection3,Failure_in_ConicalIntersection4}
Spin-flip (SF) approaches, including SFCIS\cite{Krylov01_A,Krylov01_B} and SF-TDDFT,\cite{Shao03} partially remedy this problem by accessing multiple low-spin states from a high-spin reference. Because spin-flip excitations span configurations corresponding to different electronic characters, these methods can reproduce the characteristic topology of CXs in many cases.\cite{Minezawa09, Huix10}

An alternative strategy for recovering static correlation is spin projection.\cite{Lowdin55, Mayer80, Jimenez12}
By projecting broken-symmetry determinants onto proper spin eigenstates, spin projection restores spin symmetry and improves the qualitative description of strongly correlated systems.\cite{Samanta12,Rivero13,Stein14} When combined with correlated treatments, it can approach quantitative accuracy.\cite{Schlegel86,Tsuchimochi16A,Tsuchimochi16B, Tsuchimochi19}
Within CIS-based frameworks, this concept led to time-dependent projected Hartree--Fock and spin-extended CIS (ECIS) to treat excited states.\cite{Tsuchimochi15}
Although ECIS improves descriptions of excited states in strongly correlated regimes, it remains a linear-response method based on a single reference state and is therefore not designed for a balanced multistate treatment required at CXs.\cite{Tsuchimochi15B}

The underlying difficulty in CIS, TDDFT, and related approaches can also be traced to  strong ground-state orbital bias. Orbitals optimized for the closed-shell ground state are generally ill-suited for excited states with qualitatively different electronic character such as charge-transfer states\cite{Subotnik2011, Tsuchimochi24}, core excited states,\cite{ESMF_core} and Rydberg states.\cite{Tsuchimochi26}
Near a CX, where two states differ dramatically in electronic structure and become nearly degenerate, balanced orbital relaxation also becomes essential.

This perspective motivates state-averaged orbital optimization within a CIS framework.
Recently, state-averaged formulations of orbital-optimized CIS (SACIS) and its spin-projected extension, state-averaged ECIS (SAECIS), were introduced.\cite{Tsuchimochi26} By optimizing orbitals with respect to a state-averaged objective, these methods remove preferential bias toward any single state and provide a variational description of multiple interacting states at mean-field cost. Deliberate spin-symmetry breaking followed by projection in SAECIS introduces additional variational flexibility, effectively allowing access to singlet configurations with partial double-excitation character as previously observed in ECIS.\cite{Tsuchimochi15}

While SACIS and SAECIS show promising qualitative behavior, practical applications to geometry optimization and minimum-energy CX (MECX) searches require analytical nuclear gradients. Formally, their derivation is straightforward and follows the standard Lagrangian framework.\cite{Helgaker88,Jorgensen88,Helgaker_book} Analytical derivatives for spin-projected methods have also been formulated.\cite{Schutski14, Tsuchimochi17} 
In the state-averaged framework of SACIS and SAECIS, the first derivative of the Lagrangian involves the electronic Hessian, which can  be constructed using our previous derivations.\cite{Tsuchimochi24,Tsuchimochi26}
However, redundant parametrization in these methods renders the electronic Hessian singular, leading to nonunique solutions in the associated coupled perturbed equations. 
Without careful treatment, this null space contaminates the Z-vector solution and the resulting gradients. In this work, we derive analytical nuclear gradients for SACIS and SAECIS and introduce an explicit projection procedure to eliminate null-space contributions, ensuring numerically stable and physically meaningful gradients.

As a stringent test, we investigate the well-known pyramidalization--torsion CX of ethylene, a canonical benchmark for CX topology. We then extend our analysis to a broader benchmark set of minimum-energy CXs (MECXs), comparing the performance of SACIS and SAECIS with multireference configuration interaction (MRCI) and SF-TDDFT references.
Through these analyses, we evaluate the extent to which state-averaged orbital optimization---with and without spin projection---provides a reliable and computationally efficient route to CX optimization.

This paper is organized as follows.
Section II reviews the theoretical formulation of state-averaged CIS and SAECIS and presents the derivation of their analytical nuclear gradients, including treatment of the null space in the associated coupled-perturbed equations.
Section III summarizes the computational details.
Section IV reports numerical results for the prototypical ethylene CX and a benchmark set of twelve MECXs.
Finally, Section V concludes with a summary and discussion of the implications of our findings.

\section{Theory}
\subsection{State-averaged CIS and ECIS}
Here, we briefly review the formulation of state-averaged CIS and ECIS. Our goal is to simultaneously optimize the ground state ($I=1$) and several excited states ($I=2,3,\ldots, n$) described by CIS wave functions,
\begin{align}
    |0_I\rangle = \sum_\mu c_\mu^I |\Phi_\mu\rangle
\end{align}
or, in the ECIS formulation, by applying a spin-projection operator $\hat P$,
\begin{align}
    \hat P |0_I\rangle = \sum_\mu c_\mu^I \hat P|\Phi_\mu\rangle
\end{align}
through minimization of the averaged energy of $n$ states. The determinant space $|\Phi_\mu\rangle$ is restricted to the HF-like reference determinant $|\Phi_0\rangle$ and singly excited determinants $|\Phi_{i}^a\rangle$. In the following, we focus primarily on the ECIS formulation;  the CIS case is recovered in the limit $\hat P \rightarrow 1$.

At convergence of the CI coefficients, the Hamiltonian and overlap matrices in the projected subspace are diagonal:
\begin{align}
   \langle 0_I|\hat H \hat P |0_J\rangle &= E_{0,I}\delta_{IJ}\label{eq:ortho1},\\
  \langle 0_I | \hat P |0_J\rangle &= \delta_{IJ}\label{eq:ortho2}
\end{align}
where $E_{0,I}$ denotes the energy of the properly orthonormalized state $\hat P |0_I\rangle$. 

As discussed in Ref.~[\onlinecite{Tsuchimochi26}], orbital rotations must be applied prior to spin projection, since applying projection first would not alter the symmetry components contained in the broken-symmetry state $|0_I\rangle$. 
However, because the projection operator $\hat P$ is non-unitary, the combined transformation $e^{\hat \lambda} \hat P e^{-\hat \lambda}$ does not preserve orthogonality. 
Therefore, when orbital rotations and spin projection are treated simultaneously, orthogonality between projected states is not guaranteed:
\begin{align}
    \langle 0_I|e^{{}^{\rm o}\hat \lambda} \hat P e^{-{}^{\rm o}\hat \lambda}|0_J\rangle \ne \delta_{IJ}.
\end{align}
As a result, the standard parametrization based on a state-transfer operator\cite{Helgaker_book} cannot be directly applied.

To address this issue, we employ a linear parametrization in which the $I$th ECIS state is written as
\begin{align}
|\tilde 0_I[{^{\rm o}}{\bm\lambda},{^{\rm c}}{\bm\lambda}^I]	\rangle = \hat P e^{-{}^{\rm o}\hat \lambda}\left(|0_I\rangle 	 + e^{{}^{\rm o}\hat \lambda}\hat {\cal Q}[{^{\rm o}\bm\lambda}]  e^{-{}^{\rm o}\hat \lambda}|{^{\rm c}}{\bm\lambda}^I\rangle\right) \label{eq:SA0tilde}
\end{align}
where $|{^{\rm c}}{\bm\lambda}^I\rangle$ represents the CI perturbation.
To ensure the orthogonality $\{\hat P e^{-\hat \lambda}|0_I\rangle\} \perp \{\hat P e^{-\hat \lambda}|{}^{\rm c}{\bm \lambda}^I\rangle\}$, we introduced the following generalized projection operator 
\begin{align}
        \hat {\cal Q}[{^{\rm o}\bm\lambda}] &= 1 - \sum_{IJ}^n \hat P e^{-{}^{\rm o}\hat \lambda}|0_I\rangle N^{-1}_{IJ} \langle 0_J| e^{{}^{\rm o}\hat \lambda} \hat P 
\end{align}
Even with this construction, however, the norm and mutual orthogonality of the basis states $\{|\tilde 0_I\}$ are not preserved under the perturbation of ${^{\rm o}\bm\lambda}$. Specifically, when ${^{\rm o}\bm\lambda} \ne {\bf 0}$,
\begin{align}
    \langle \tilde 0_I | \hat P |\tilde 0_J\rangle \ne \delta_{IJ}
\end{align} 
owing to the first term in Eq.~(\ref{eq:SA0tilde}). 
This indicates that orbital perturbations in $|\tilde 0_J\rangle$ implicitly affect the energy of other states $|\tilde 0_I\rangle$, even though the latter does not explicitly depend on the former, and vice versa. Consequently, the naive average of state-specific expectation values, $\frac{1}{n}\sum_I \langle \tilde 0_I|\hat H |\tilde 0_I\rangle/\langle \tilde 0_I | \tilde 0_I\rangle$, is not an appropriate objective for state-averaged ECIS. In other words, the energy expectation value does not correspond to the properly orthogonalized energies $E_I$, and the coupling between states must be treated explicitly through the Hamiltonian and overlap matrices,
\begin{align}
    {\mathcal H}_{IJ}[{^{\rm o}}{\bm\lambda},{^{\rm c}}{\bm\lambda}^I,{^{\rm c}}{\bm\lambda}^J] &= \langle \tilde 0_I[{^{\rm o}}{\bm\lambda},{^{\rm c}}{\bm\lambda}^I] | \hat H |\tilde 0_J[{^{\rm o}}{\bm\lambda},{^{\rm c}}{\bm\lambda}^J] \rangle 
    \\ 
 {\mathcal N}_{IJ}[{^{\rm o}}{\bm\lambda},{^{\rm c}}{\bm\lambda}^I,{^{\rm c}}{\bm\lambda}^J] &= \langle \tilde 0[{^{\rm o}}{\bm\lambda},{^{\rm c}}{\bm\lambda}^I] |\tilde 0[{^{\rm o}}{\bm\lambda},{^{\rm c}}{\bm\lambda}^J] \rangle
  \end{align}
Note that in standard (non-projected) methods such as SACIS, orthogonality is preserved, i.e., ${\mathcal N}_{IJ} = \delta_{IJ}$, due to the unitarity of $e^{-\hat \lambda}$. The loss of orthogonality in spin-projected methods arises from the non-unitarity of  $\hat P$.

The energies $E_I$ are therefore defined only after explicit orthogonalization by solving the generalized eigenvalue problem
\begin{align}
    {\bm{\mathcal H}}{\bf V} = {\bm{\mathcal N}}{\bf V}{\bf E} \label{eq:HVSVE}
\end{align}
where $\bm{\mathcal H}$ and $\bm{\mathcal N}$ are the Hamiltonian and overlap matrices in the subspace $|\tilde 0_I\rangle;\;(I=1,\cdots,n)$, {\bf V} contains the eigenvectors, and {\bf E} is the diagonal matrix of eigenvalues $E_I$. This formulation allows the weighted state-averaged energy to be written as
\begin{align}
    E_{\rm ave}[{\bf W}] = {\rm Tr}[{\bf W}{\bf V}^\dag {\bm {\mathcal N}}^{-1}{\bm {\mathcal H}} {\bf V}]
\end{align}
where the diagonal weight matrix $\mathbf W$ contains elements $w_I$. For equal weighting ($w_I = 1/n$), the expression simplifies to 
\begin{align}
    E_{\rm ave} = \frac{{\rm Tr}[{\bm{\mathcal {N}}}^{-1}{\bm {\mathcal H}}]}{n}
\end{align}
This way, $E_{\rm ave}$ becomes a function of the parameter set $
    \bm\lambda = \begin{pmatrix}
        {^{o}\bm\lambda} & {^{c}\bm\lambda}^1 & {^{c}\bm\lambda}^2 &\cdots&{^{c}\bm\lambda}^{n})
    \end{pmatrix}
        ^\top
    $.
In the limit $n=1$, this expression reduces to the single-state ECIS (SS-ECIS) formulation.

\subsection{SAECIS Lagrangian and its derivative}
In the single-root, state-specific ECIS, both the orbitals and the CI coefficients are optimized with respect to a single-state energy. The fully variational feature of the energy functional of SS-ECIS allows for the straightforward derivation of its analytical derivative. In fact, the analytical gradient for ECIS can be obtained directly from the CISD formalism\cite{Tsuchimochi17} by omitting the double excitation coefficients.

In contrast, in SAECIS, the averaged energy is minimized, and the individual energies $E_I$ are not variational with respect to orbital rotations $^{\rm o}{\bm \lambda}$. They are stationary only with respect to the CI coefficients $^{\rm c}{\bm \lambda}^I$. Consequently, evaluation of energy gradients for individual states requires construction of an appropriate Lagrangian to enforce the necessary stationarity conditions.

The SAECIS Lagrangian for the $I$th state can be constructed as
\begin{align}
    {\mathscr L}_I & = E_I + \sum_{ai} {^{\rm o}}z_{ai}^I \frac{\partial E_{\rm ave}}{\partial\; {^{\rm o}\lambda_{ai}}}
    + \sum_K \sum_\mu {^{\rm c}}z_{\mu}^{IK} \frac{\partial E_{\rm ave}}{\partial\; {^{\rm c}}\lambda_\mu^K}\label{eq:L}
\end{align}
where ${^{\rm o}}z_{ai}^I$ and ${^{\rm c}}z_\mu^{IK}$ are the Lagrange multipliers to be determined. The orbital gradient of the averaged energy is simply the average of the orbital gradients of the individual state energies $E_I$. In other words,
\begin{align}
    \frac{\partial E_{\rm ave}}{\partial {}^{\rm o}\lambda_{ai}} = \frac{1}{n}\sum_K^n {}^{\rm o} g^K_{ai} 
\end{align}
where 
\begin{align}
        {}^{\rm o}g^K_{ai} &= \frac{\partial E_K}{\partial {}^{\rm o}\lambda_{ai}}\br
        &= \langle 0_K|(a^\dagger_a a_i- a^\dagger_i a_a) (\hat H - E_{K})\hat P |0_K\rangle + h.c.
\end{align}
Similarly, the CI gradient is
\begin{align}
        \frac{\partial E_{\rm ave}}{\partial {}^{\rm c}\lambda_{\mu}} = \frac{1}{n}\sum_K^n {}^{\rm c} g^K_{\mu}
\end{align}
with
\begin{align}
        {}^{\rm c}g^K_{\mu} &= \frac{\partial E_I}{\partial {}^{\rm c}\lambda_{\mu}}\br
        &= \langle \Phi_\mu|(\hat H - E_{K})\hat P |0_K\rangle + h.c.
        \end{align}
We require ${\mathscr L}_I$ to be stationary with respect to all variational parameters (${^{\rm o}}\lambda_{ai}, {^{\rm c}}\lambda_\mu^I, {^{\rm o}}z_{ai}^I, {^{\rm c}}z_\mu^{IK}$). This leads to the following coupled perturbed (CP) SACIS equations:
\begin{align}
\frac{\partial {\mathscr L}_I}{\partial \kappa_{ai}} &= \frac{\partial E_I}{\partial \;{^{\rm o}}\lambda_{ai}} + \sum_{bj} \frac{\partial^2 E_{\rm ave}} {\partial \;{^{\rm o}}\lambda_{ai}\partial\; {^{\rm o}}\lambda_{bj}} {^{\rm o}}z_{bj}^I 
\br 
&+\sum_K \sum_\nu  \frac{\partial^2 E_{\rm ave}}{\partial\; {^{\rm o}}\lambda_{ai} \partial \;{^{\rm c}}\lambda_\nu^K} {^{\rm c}}z_\nu^{IK} = 0
\\
\frac{\partial {\mathscr L}_I}{\partial c_{\mu}^J} &=
\sum_{bj} \frac{\partial^2 E_{\rm ave}} {\partial \;{^{\rm c}}\lambda_\mu^J\partial \;{^{\rm o}}\lambda_{bj}} {^{\rm o}}z_{bj} ^I
+\sum_{K}\sum_\mu  \frac{\partial^2 E_{\rm ave}}{\partial  \;{^{\rm c}}\lambda_\mu^J \partial \;{^{\rm c}}\lambda_\nu^K} {^{\rm c}}z_\nu^{IK}  = 0
\end{align}

Recognizing that the quantities appearing above  correspond to the gradient and Hessian of the averaged energy, the CP-SACIS equations can be recast as 
\begin{align}
    {\bf H}{\bf z} + {\bf g} = {0}\label{eq:Hz+g=0}
\end{align}
where
\begin{align}
      {\bf H} &= \begin{pmatrix}
         {^{\rm oo}}{\bf H} &  {^{\rm oc}}{\bf H}^1 &  {^{\rm oc}}{\bf H}^2 & \cdots  &{^{\rm oc}}{\bf H}^{n}\\
          {^{\rm co}}{\bf H}^1 &  {^{\rm cc}}{\bf H}^{11} & {\bf 0} & \cdots & {\bf 0}\\
       {^{\rm co}}{\bf H}^2  & {\bf 0} &  {^{\rm cc}}{\bf H}^{22}& \cdots & {\bf 0}\\
        \vdots & \vdots & \vdots &\ddots & \vdots\\
    {^{\rm co}}{\bf H}^{n}  & {\bf 0}  & {\bf 0}&\cdots & {^{\rm cc}}{\bf H}^{nn} \\
    \end{pmatrix}\\
    {\bf g} &= \begin{pmatrix}
        {}^{\rm o}{\bf g}^I&
        {\bf 0}& \cdots&
        {\bf 0}
    \end{pmatrix}^\top\\
        {\bf z} &= \begin{pmatrix}
        {}^{\rm o}{\bf z}^I&
        {}^{\rm c}{\bf z}^{I1}& 
        \cdots&
        {}^{\rm c}{\bf z}^{In}
    \end{pmatrix}^\top
\end{align}
The Hessian blocks are defined as
\begin{align}
    {}^{\rm oo}H_{ai, bj} &= \frac{\partial^2 E_{\rm ave}}{\partial {}^{\rm o}\lambda_{ai} \partial {}^{\rm o}\lambda_{bj}}\\
    {}^{\rm oc}H^I_{ai,\mu} &= \frac{\partial^2 E_{\rm ave}}{\partial {}^{\rm o}\lambda_{ai} \partial {}^{\rm c}\lambda_{\mu}^I} = {}^{\rm co}H^I_{\mu,ai}\\
    {}^{\rm cc}H^{IJ}_{\mu\nu} &= \frac{\partial^2 E_{\rm ave}}{\partial {}^{\rm c}\lambda_{\mu}^I \partial {}^{\rm c}\lambda_{\nu}^I } \delta_{IJ}
\end{align}
whose explicit working expressions are given in Refs.~\cite{Tsuchimochi24, Tsuchimochi26} Using these definitions, the CP-SACIS equations can be written explicitly as
\begin{align}
        \sum_{bj} {^{\rm oo}}H_{ai, bj} {^{\rm o}}z_{bj}^I + \sum_{J, \nu}  {^{\rm oc}}H^J_{ai, \nu}{^{\rm c}}z^{IJ}_\nu &= - {^{\rm o}}g^I_{ai} \label{eq:Zvec} \\
     \sum_{ai}{^{\rm co}}H^J_{\mu, ai} {^{\rm o}}z^I_{ai} + \sum_{\nu} {^{\rm cc}}H^{JJ}_{\mu, \nu}{^{\rm c}}z^{IJ}_{\nu} &=0 \;\;\;\;\forall \;J
\end{align}

It is important to note that the Hessian matrix is generally singular due to redundant parametrization in the CI space. As a result, multiple solutions ${\bf z}$ may satisfy the above linear equations. This issue will be addressed in Sec.~\ref{sec:null}.

\subsection{Nuclear gradients} \label{sec:nuc}
The analytical energy derivative is central to the evaluation of nuclear gradients, required for efficient geometry optimization. In general, the derivative of the $I$th state energy with respect to a nuclear perturbation $x$, $E_I^x = dE_I/dx$, can be written as
\begin{align}
    E_I^{x} = \sum_{\mu\nu} h_{\mu\nu}^x P_{\mu\nu}^{I, \rm rel} + \frac{1}{4} \sum_{\mu\nu\lambda\sigma}\langle \mu\nu||\lambda\sigma \rangle ^x P^{I, \rm rel} _{\nu\mu, \sigma\lambda} - \sum_{\mu\nu}S^x_{\mu\nu} W^I_{\nu\mu}\label{eq:EIx}
\end{align}
where $h_{\mu\nu}^x$, $\langle \mu\nu||\lambda\sigma\rangle^x$, and $S_{\mu\nu}^x$ denote derivatives of the one- and two-electron integrals and the overlap matrix in the AO basis, with respect to $x$. ${\bf P}^{I, \rm rel}$ and ${\bf W}^I$ are the so-called relaxed density matrix and energy-weighted density matrix, respectively. Because the Lagrangian ${\cal L}_I$ is fully stationary at the optimized wave function and equals $E_I$, application of the chain rule gives $ E_I^{x} = {\mathcal L}_I^{x} = {\mathcal L}_I^{(x)}$, where the superscript $(x)$ denotes a partial derivative taken with orbitals and CI coefficients held fixed. Using Eq.(\ref{eq:L}), this yields
\begin{align}
    {\mathscr L}_I^{(x)} = E_I^{(x)} + \frac{1}{n}\sum_K \sum_{ai} {^{\rm o}}z_{ai}^I {^{\rm o}}g^{K (x)}_{ai} + \frac{1}{n}\sum_K \sum_\mu {^{\rm c}}z_{\mu}^{IK}  {^{\rm c}}g^{K (x)}_{\mu} 
\end{align}
Our objective here is to recast this expression into the standard form of Eq.~(\ref{eq:EIx}). 

Following previous work,\cite{Tsuchimochi17} the above equation can be rewritten as
\begin{align}
    {\mathscr L}_I^{(x)} = {\mathscr L}_I^{(\bar x)}  
 - \sum_g  \frac{\partial {\mathscr L}_{I}}{\partial ( {\bm{\mathcal L}}_g{\bf C}^\dagger)_{r\mu} }({\bm{\mathcal L}}_g{\bf C}^\dag{\bf S}^x {\bf S}^{-1})_{r\mu}
 \label{eq:LIx_2}
\end{align}
by using the chain rule. Here, $(\bar x)$ indicates that only the explicit integral derivatives $h_{\mu\nu}^x$ and $\langle \mu\nu||\lambda\sigma\rangle^x$ are included, i.e., the Pulay contribution arising from the nuclear dependence of the basis functions is excluded. The second term in Eq.~(\ref{eq:LIx_2}) corresponds to the Pulay force and represents the contraction between the energy-weighted density matrix ${\bf W}^I$ and the overlap derivative ${\bf S}^x$, as in the final term of Eq.~(\ref{eq:EIx}). Here, $g$ labels the numerical grid points used in spin projection and ${\bm{\mathcal L}}_g$ is an intermediate matrix at grid point $g$ arising naturally from the non-orthogonal Wick theorem.\cite{Tsuchimochi16B, Tsuchimochi17} Eq. (\ref{eq:LIx_2}) holds for any general expectation values with grid-based spin-projection. For standard (non-projected) CIS, ${\bm{\mathcal L}}_g$ reduces to the identity matrix, and Eq.~(\ref{eq:LIx_2}) simplifies to
\begin{align}
    {\mathscr L}_I^{(x)} = {\mathscr L}_I^{(\bar x)}  
 - \frac{\partial {\mathscr L}_{I}}{\partial  C^*_{r\mu} }({\bf C}^\dag{\bf S}^x {\bf S}^{-1})_{r\mu}
\end{align}

The relaxed density matrix and energy-weighted density matrix can then be expressed as
\begin{align}
    {\bf P}^{I,\rm rel} &= {\bf P}^I + {^{\rm o}}{\bf P}^I + {^{\rm c}}{\bf P}^I\\
    {\bf W}^{I} &= \tilde {\bf W}^{I} + {^{\rm o}}{\bf W}^I + {^{\rm c}}{\bf W}^I
\end{align}
Further details, including their explicit forms, are  provided in the Supplemental Information.

\subsection{Treatment of redundancies}\label{sec:null}
Because of the redundant parametrization introduced in the CI space, the Hessian matrix possesses a null space in both SACIS and SAECIS. As a consequence, the corresponding CP equations admit non-unique solutions. Specifically, the linear system Eq.~(\ref{eq:Hz+g=0}) has infinitely many solutions, since any vector in the null space of {\bf H} can be added to a particular solution. Thus, the general solution can be written as
\begin{align}
    \tilde{\bf z} = {\bf z} + \sum_{k}^{n_{\rm red}}\alpha_k {\bf y}_k
\end{align}
with ${\bf z}$ being the unique ``pure'' solution. Here, ${\bf y}_k$ denote the $n_{\rm red}$ null vectors (normalized for convenience) that satisfy
\begin{align}
    {\bf H}{\bf y}_k = {\bf 0}
\end{align}
and $\alpha_k\in\mathbb{C}$ are arbitrary coefficients.

In practice, iterative solvers such as conjugate gradient often converge to solutions $\tilde{\bf z}$ containing large components along these null directions. Although such components formally satisfy Eq.~(\ref{eq:Hz+g=0}), they contaminate the relaxed density matrix and lead to unphysical contributions to the nuclear gradients within the SA formulation (see Appendix). It is therefore essential to remove the null space components. While null-space projection can be carried out by repeatedly applying the metric of the Hessian to ${\tilde {\bf z}}$,\cite{Tsuchimochi18} for CIS the Hessian structure is simpler, and the null space can be identified and projected out explicitly. 

At a stationary point, the null space of the CIS/ECIS Hessian originates exclusively from the CI block, ${}^{\rm cc}{\bf H}^I$. Indeed, the state vectors $|0_K\rangle$ span this null space, independent of $I$, as
\begin{align}
    \sum_\nu {}^{\rm cc}{H}^I _{\mu\nu} c_\nu^K = 0 \label{eq:vanish1}
\end{align}
where $c_\nu^K$ are the CIS coefficients of $|0_K\rangle$.
The corresponding orbital-CI coupling also vanishes,
\begin{align}
    \sum_\nu {}^{\rm oc}{H}^I _{\mu\nu} c_\nu^K = 0
    \label{eq:vanish2}
\end{align}
as shown in the Supplemental Information. By construction, the orbital block itself contains no redundancy since  only occupied-virtual rotations are allowed. 

Eqs.~(\ref{eq:vanish1})--(\ref{eq:vanish2}) follow from our choice of CI basis, $\left\{\left(1 - \hat P|0_K\rangle \langle 0_K|\hat P\right) \hat P|\Phi_\mu\rangle\right\}$, which removes all averaged ECIS states from the projected determinant space, $\{\hat P |\Phi_\mu\rangle\}$. With excitation rank restricted to singles, these projected determinants generally contain no additional internal redundancies. Consequently, the CI space contains exactly $n$ redundant directions per state block. Physically, this redundancy reflects the fact that the broken-symmetry singly excited determinants $|\Phi_i^a\rangle$ and $|\Phi_{\bar i}^{\bar a}\rangle$ contain distinct singlet components. 
The only exception occurs when subsets of $\alpha$ and $\beta$ orbitals simultaneously span identical subspaces within the occupied and virtual manifolds, thereby preserving partial spin symmetry.\cite{CUHF1,CUHF2}
In such cases, the configuration $|\Phi_i^a\rangle - |\Phi_{\bar i}^{\bar a}\rangle$ is purely triplet and destroyed after singlet projection, resulting in a null basis (the extreme case being spin projection applied to spin-restricted CIS). Having said that, this does not happen when the ECIS wave function is fully variationally optimized, unless constraints are deliberately imposed, as in constrained UHF.\cite{CUHF1,CUHF2}
In general, spin projection converts these triplet-like CIS configurations into additional singlet states through the projection operator $\hat P$, often corresponding to states with doubly excited electronic character.
For spin-restricted SACIS, one may work directly in the reduced space containing only singlet couplings without triplet couplings. In this case, the number of redundancies is again $n$, for the same reasons outlined above.

Taken together, these results show that the CI coefficient vectors $\{{\bf c}^K\}$ span the null space of each CI block, forming $\{{\bf y}_k\}$ with $n_{\rm red} = n^2$, unless {\bf H} contains additional accidental zero eigenvalues. We therefore construct the null-space projector ${\bm{\mathcal P}}$ using ${\bf Y} = \begin{pmatrix}
    {\bf c}^1 & {\bf c}^2 & \cdots &  {\bf c}^n
\end{pmatrix}$. 
In SACIS, all CI vectors are mutually orthonormal, resulting in ${\bf Y}^\dagger {\bf Y} = {\bf I}$. By contrast, in SAECIS, the CI vectors are generically neither normalized nor orthogonal; strict orthonormality is achieved only through the projection operator. Consequently, they cannot by themselves serve as a proper basis for projecting out the null space. To address this, we form the following projector in each CI space, ${\bm{\mathcal P}}$, by orthonormalizing {\bf Y},
\begin{align}
 {\bm{\mathcal P}} = {\bf I} - {\bf Y}({\bf Y}^\dagger {\bf Y})^{-1}{\bf Y}^\dagger
\end{align}
We then apply this projector to each CI block of the contaminated Z-vector:
\begin{align}
    {\bf z} = \begin{pmatrix}
        {\bf I} & {\bf 0} & \cdots & {\bf 0} \\
        {\bf 0} & {\bm{ \mathcal P}} & \cdots & {\bf 0}\\
        \vdots & \vdots &\ddots & {\bf 0}\\
        {\bf 0} & {\bf 0} & \cdots & {\bm{\mathcal P}}\\
    \end{pmatrix}
    \begin{pmatrix}
        {}^{\rm o} {\bf z}\\
        {}^{\rm c}\tilde {\bf z}^1\\
        \vdots\\
        {}^{\rm c}\tilde {\bf z}^n
    \end{pmatrix}
\end{align}
This procedure is needed only once upon convergence of $\tilde {\bf z}$ and effectively eliminates the redundant components from the Z-vector, ensuring numerical stability in subsequent computations.

\begin{table*}[t]
    \centering
    \tabcolsep = 6pt
    \caption{Optimized geometries of the ground state ($S_0$) of ethylene computed with the 6-31+G** basis set.}\label{tb:geom}
    \begin{tabular}{c|llllll}
    \hline\hline
         & $R_{\rm CC}$ & $R_{\rm CH}$ & $\angle{\rm HCH}$ &  Point-Group \\
         \hline
         SACIS & 1.322 & 1.080/1.079 & 116.7/115.7   & $C_{2v}$ \\
         SAECIS &1.334 & 1.087 &  116.1 &$D_{2h}$ \\
         B3LYP & 1.334 &1.087 & 116.5  &$D_{2h}$  \\
         SF-TDDFT& 1.330 & 1.077 & 116.7  &$D_{2h}$ \\
          SA-CASSCF& 1.324  &  1.076 & 117.1 & $D_{2h}$ \\
         Exptl. & 1.339 & 1.086 & 117.6 & $D_{2h}$\\
         \hline\hline
    \end{tabular}
\end{table*}

\section{Computational details}

SACIS and SAECIS, together with their analytical nuclear gradients, were implemented in the {\sc Gellan} suite of programs,\cite{Gellan} and all calculations were performed with this code except for SA-CASSCF and SF-TDDFT, which were carried out with OpenMolcas\cite{openmolcas} and ORCA\cite{orca}, respectively. All calculations used the 6-31+G** basis set. For SF-TDDFT, we adopted the BHHLYP functional.\cite{B3LYP}

State-averaged calculations were performed for the two lowest singlet states, $S_0$ and $S_1$. CIS variants were computed in a spin-restricted framework, whereas ECIS variants were performed in a spin-unrestricted framework followed by spin projection to recover the singlet state. Throughout all systems, we have employed an active space of (2e, 2o) for SA-CASSCF. Point-group symmetry was not imposed throughout this work.
 
In the present applications, no problematic instability of the target state-averaged solution was observed for SACIS during the MECX optimizations and PES scans considered here. For SAECIS, however, some optimization paths, such as that for butadiene, occasionally exhibited a temporary transition to an alternative state-averaged solution that was slightly more stable but had a larger energy gap. Upon further optimization toward the crossing region, the procedure returned to the branch with the small gap that continuously leads to the MECX. This behavior suggests the coexistence and crossing of distinct self-consistent state-averaged solutions in the broader nonlinear variational space of SAECIS, rather than simple root flipping in the usual sense. The reported SAECIS MECX geometry for butadiene corresponds to this latter branch.

\section{Results and discussions}
\subsection{Ethylene}
We first examine the potential energy surfaces (PESs) of the ground and excited states of ethylene and assess the ability of each method to describe CXs. 

The optimized geometries of the ground state ($S_0$) are summarized in Table~\ref{tb:geom}. Most methods correctly reproduce the experimentally observed $D_{2h}$ symmetry. Only SACIS predicts a lower $C_{2v}$ symmetry, yielding two inequivalent $R_{\rm CH}$ bond lengths and $\angle{\rm HCH}$ angles. This symmetry breaking originates from the underlying wave function.

In SACIS, the parametrization is restricted to a Hartree--Fock-like reference determinant and its single excitations. Since the wave function is variationally optimized within this limited configuration space, it can lower its energy by breaking spatial symmetry to gain additional correlation energy. This behavior is analogous to the well-known spin-symmetry breaking in Hartree--Fock theory and to the less frequently discussed spin-preserving spatial symmetry breaking observed in certain HF solutions.\cite{Li09} Nevertheless, the deviation from $D_{2h}$ symmetry in SACIS is relatively minor, which is also supported by the negligible dipole moment of 0.003 Debye.  Although SA-CASSCF preserves the $D_{2h}$ symmetry, its optimized geometry is close to that of SACIS. In particular, both methods predict a slightly shorter $R_{\rm CC}$ than the other methods considered here. When spin projection is incorporated, SAECIS restores the higher-symmetry $D_{2h}$ minimum, and the optimized geometry closely resembles that obtained with B3LYP,\cite{B3LYP}  which is in good agreement with experimental values.

\begin{figure}
    \centering
    \includegraphics[width=0.3\linewidth]{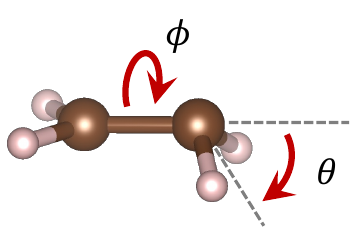}
    \caption{Torsion and pyramidalization angles of C$_2$H$_4$.}
    \label{fig:C2H4}
\end{figure}

The description of the $S_1$ state is more challenging and chemically richer. A simple theoretical consideration suggests a $D_{2d}$ structure in which the two HCH planes are perpendicular to each other, the ionic configurations (C$^+$C$^-$ and C$^-$C$^+$) are degenerate, and their coupling vanishes by symmetry. The resulting state is nonpolar and exhibits diradical character. However, this high-symmetry configuration is unstable. A slight structural distortion—specifically, pyramidalization of one of the methylene groups, characterized by the angle $\theta$ in Figure~\ref{fig:C2H4}—breaks the $D_{2d}$ symmetry and induces strong mixing between the two zwitterionic singlet configurations. This mixing localizes the $\pi$-electron pair on the pyramidalized carbon atom.\cite{Bonaci75, Brooks79, Angeli09, Filatov21, Saade24} 
Thus, the deviation of $\theta$ from $0^\circ$ represents energetic stabilization via charge separation, which can be interpreted as a manifestation of the pseudo-Jahn-Teller effect. 

In our calculations, during optimization of the $S_1$ state, several methods experienced the root-flipping issue where $S_0$ and $S_1$ become degenerate. In these cases, no stationary point satisfying $dE/dR = 0$ could be located on the $S_1$ surface. At the same time, this behavior also indicates that the method is capable of reaching the CX, demonstrating at least qualitative CX expressibility.
However, the detailed topology of the PES around the CX is not immediately evident, particularly for SACIS and SAECIS. We therefore examine the PESs explicitly for each method.

\begin{figure*}
    \includegraphics[width=150mm]{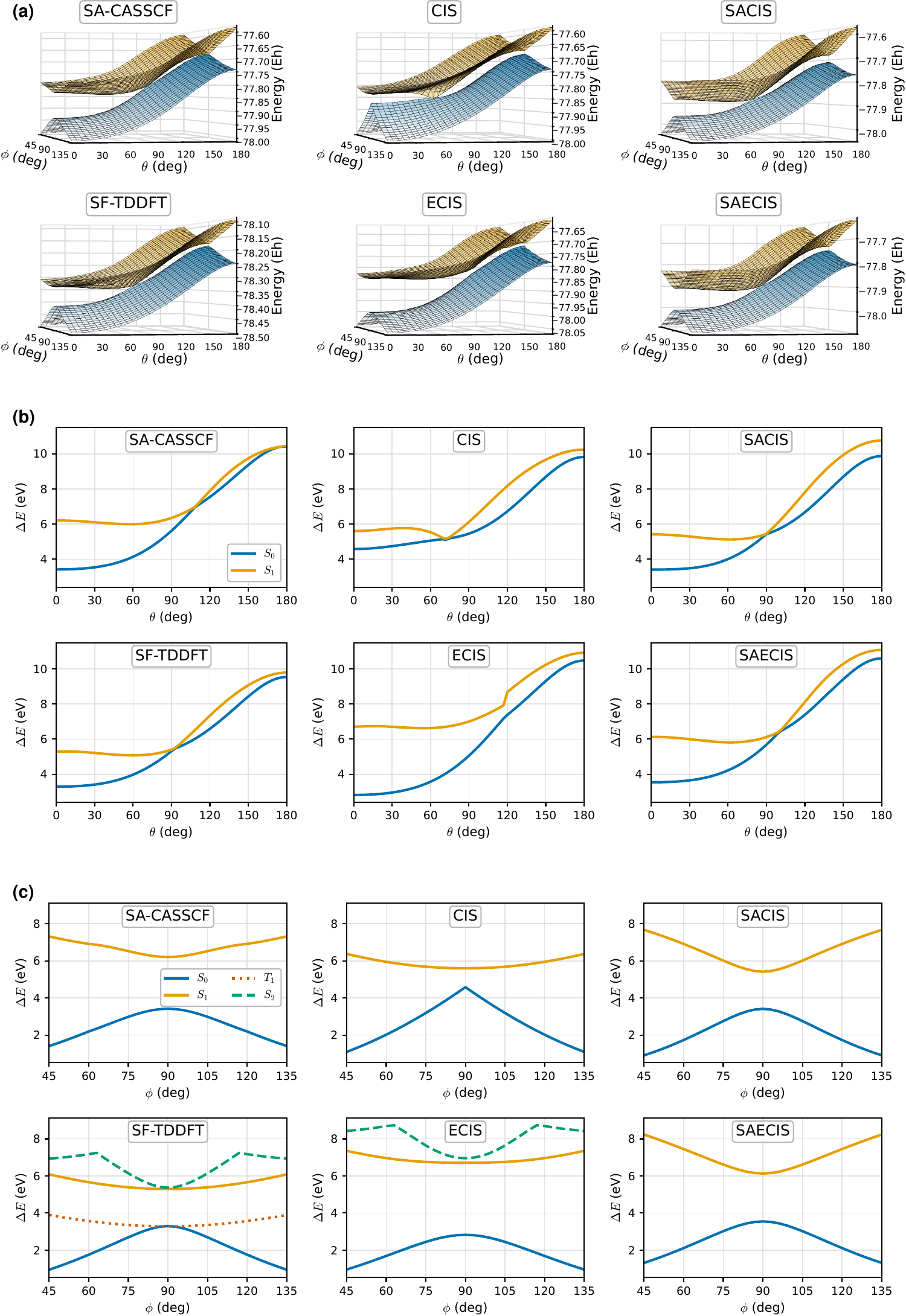}
    \caption{Potential energy surfaces of ethylene as a function of bond torsion and pyramidalization angles $\phi$ and $\theta$.  All other internal coordinates were fixed at the planar ground-state geometry optimized at the B3LYP/6-31+G** level. (a) Description of conical intersection. (b) $S_0, S_1$ energy tomography along $\phi$ at $\theta = 0^\circ$. (c) $S_0, S_1$ energy tomography along $\theta$ at $\phi$ $= 90^\circ$.}\label{fig:PES}
\end{figure*}

To this end, we employ twisted-pyramidalized (tw-pyr) ethylene (Fig.~\ref{fig:C2H4}) as a prototypical test case, using the C-C torsional angle $\phi$ and the C-CH$_2$ pyramidalization angle $\theta$ as nuclear coordinates. All scans start from the planar ground-state geometry ($\phi=\theta=0^\circ$), optimized at the B3LYP/6-31+G** level (Table~\ref{tb:geom}), and only $\phi$ and $\theta$ are varied, with all other internal coordinates fixed.
Figure~\ref{fig:PES}(a) presents the PESs of the ground ($S_0$) and  first excited ($S_1$) states computed with various CIS-based approaches. Throughout the scan range considered here, the $S_1$ state retains $\pi$-$\pi^\ast$ character.  As a reference method, SA-CASSCF(2e,2o) provides the minimal multireference description for this system and exhibits two distinct CXs, which are most clearly identified in the $\theta$-scan at fixed $\phi=90^\circ$ shown in Fig.~\ref{fig:PES}(b). In contrast, SF-TDDFT with the BHHLYP functional\cite{B3LYP} yields only a single CX. This qualitative difference likely arises from the inclusion of dynamical correlation in SF-TDDFT, which alters the topology of the PES.

As is well known, CIS fails to reproduce the correct CX topology. In Fig.~\ref{fig:PES}(b), the CIS surfaces display a clear discontinuity near the putative intersection region. This deficiency can be traced to the single-reference nature of the Hartree-Fock ground state, which lacks static correlation. The absence of static correlation is further evident from Fig.~\ref{fig:PES}(c), which shows the PES cut at $\theta=0^\circ$ (pure torsion): the $S_0$ curve exhibits a pronounced cusp at $\phi=90^\circ$. The spin-projected variant ECIS maintains a finite $S_0$-$S_1$ energy gap throughout the scan, and the two surfaces never intersect. Consequently, neither CIS nor ECIS can describe CXs.

Remarkably, SACIS provides a qualitatively reasonable description of the CX, despite being restricted to single excitations. This behavior can be attributed to the strong orbital relaxation inherent in the fully variational SACIS framework. The optimization effectively induces charge-transfer-like localization of the orbitals, thereby compensating for the nominal lack of higher excitations. As shown in Fig.~\ref{fig:PES}(c), SACIS correctly captures the degeneracy at $\theta=0^\circ$ and $\phi=90^\circ$, indicating recovery of the static correlation associated with the twisted geometry. This behavior can be interpreted as performing CIS on a spatial-symmetry-broken reference resembling one of the two zwitterionic configurations, $(\mathrm{CH}_2)^+=(\mathrm{CH}_2)^-$, which restores the near-degeneracy between bonding and antibonding orbitals as a single excitation from $(\mathrm{CH}_2)^-$ to $(\mathrm{CH}_2)^+$. Thus, although SACIS is formally limited to single excitations, the combination of orbital optimization and spatial symmetry breaking enables it to mimic key features of multireference character. The same localization mechanism is likely to be responsible for the symmetry breaking observed in the ground state. In the present scans, this symmetry breaking does not appear to introduce pathological discontinuities or spurious artifacts in the PES. Rather,  this multireference character is key to obtaining smooth curves throughout the PES and the correct dimensionality of CX in SACIS. 

As seen in Figs.~\ref{fig:PES}(a) and (b), the overall topography of the SACIS PES more closely resembles that of SF-TDDFT than that of CASSCF. This difference can be attributed to correlation effects beyond the minimal CASSCF(2e,2o) active space. In CASSCF, the configuration space is restricted by the choice of active orbitals, whereas in SACIS it is governed by excitation rank combined with full orbital optimization. This flexibility allows SACIS to incorporate correlation effects that are absent in the small active space treatment.

\begin{figure}
    \centering
    \includegraphics[width=0.5\linewidth]{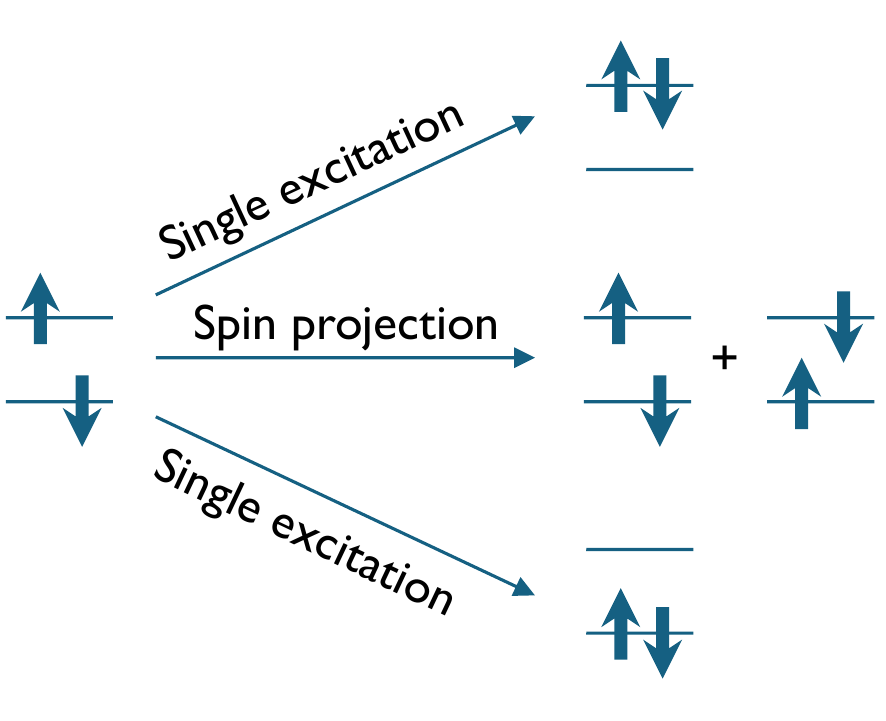}
    \caption{Configuration space accessible within ECIS.}
    \label{fig:orb}
\end{figure}

SAECIS closely mirrors SACIS across all scans, indicating that spin projection does not qualitatively modify the CX topology in this system. This suggests that SACIS already provides an adequate zeroth-order description of the CX. A distinct advantage of SAECIS, however, lies in its improved description of the higher-lying $S_2$ state directly above $S_1$, when $n=3$. Although not shown explicitly, this is evident from the ECIS results in Fig.~\ref{fig:PES}(c), where the $S_2$ surface is described reasonably well, even though it is expected to have substantial doubly excited character. The physical mechanism by which ECIS achieves this within a nominally single-excitation framework was elucidated in Ref.[\onlinecite{Tsuchimochi15}], but it can also be understood from Figure \ref{fig:orb}. With state-averaged orbital optimization, the reference state $|\Phi_0\rangle$ may become a broken-symmetry non-aufbau state. Single excitations from such a reference generate configurations corresponding to closed-shell and doubly excited configurations, while spin projection yields the open-shell singlet configuration. This qualitatively accurate treatment becomes particularly important when three states are included in the state-averaging procedure. In that case, a balanced description of $S_0$, $S_1$, and $S_2$ is essential to ensure a physically consistent multistate representation. By contrast, SACIS has the intrinsic limitation of a single-excitation ansatz, and therefore the $S_2$ state is inaccessible.

Finally, although these scans indicate that the SACIS and SAECIS surfaces remain locally smooth near the MECX, a more rigorous branching-plane analysis based on the gradient-difference and interstate-coupling vectors would be desirable. The latter is not yet available in the present implementation, even at the finite-difference level, because it would require additional implementation for robust evaluation and tracking of the intersecting states at displaced geometries. The present $\theta$- and $\phi$-dependent scans are therefore intended as a practical assessment of the local smoothness and qualitative topology of the SACIS and SAECIS surfaces near the MECX.

\begin{figure*}
    \centering
    \includegraphics[width=1\linewidth]{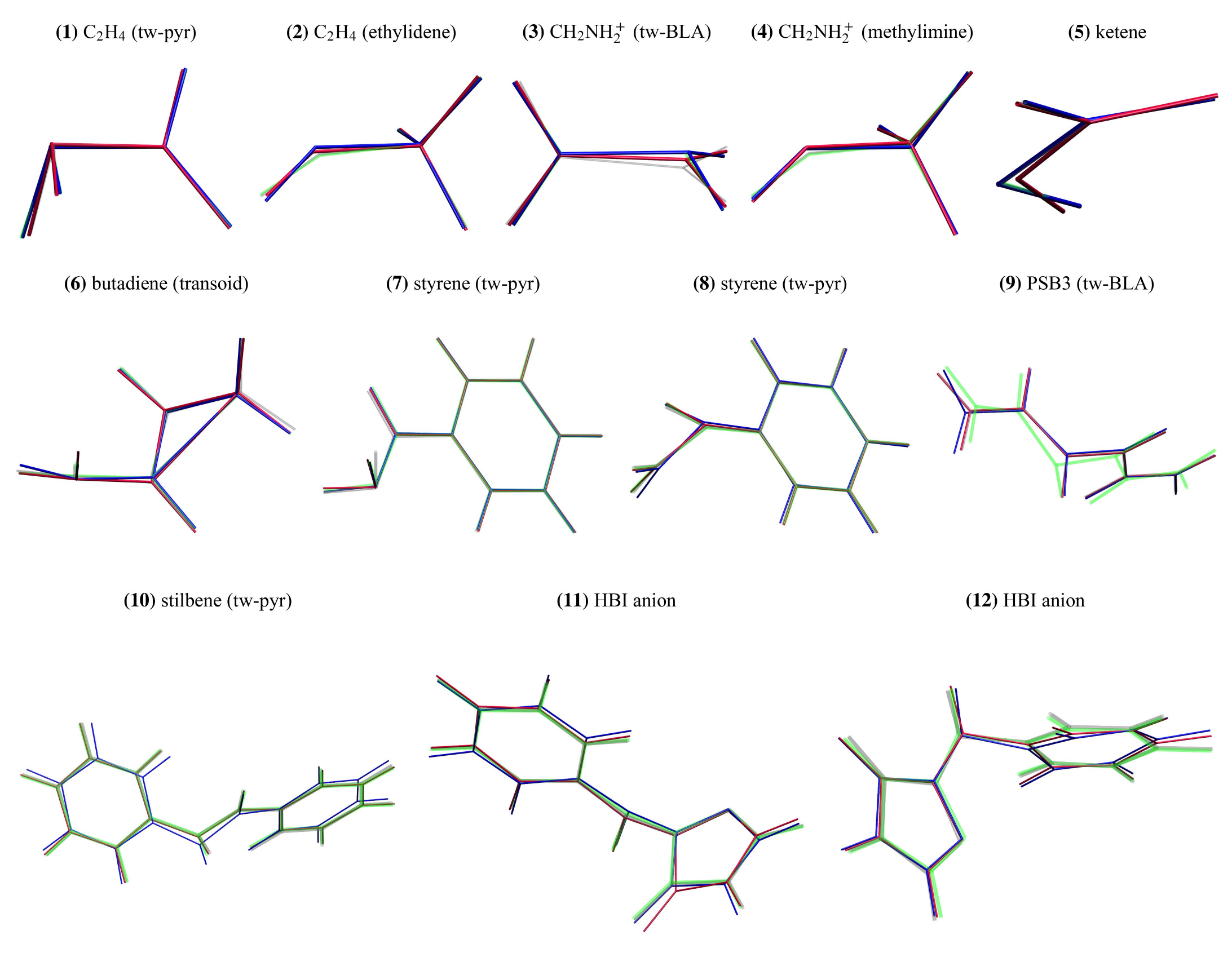}
    \caption{Comparison of MECX structures. Gray: MRCI, Green: SF-TDDFT, Blue: SACIS, Red: SAECIS.}
    \label{fig:CX}
\end{figure*}

\subsection{Minimum Energy Conical Intersection}

Having established that both SACIS and SAECIS provide qualitatively correct CX topologies, we now investigate the MECXs between $S_0$ and $S_1$.

A rigorous optimization of an MECX requires the gradient difference and derivative coupling vectors that define the branching plane. While the former is readily available once the energy gradient is available, the latter requires additional derivation and implementation. Instead of computing them explicitly, we follow the penalty-function strategy of Levine and Martinez,\cite{Levine08} which recasts the MECX search as the minimization of a single scalar objective function,
\begin{align}
F[{\bf R}] = \frac{E_{S_0}[{\bf R}] + E_{S_1}[{\bf R}]}{2}+
\sigma \frac{\left(E_{S_1}[{\bf R}] - E_{S_0}[{\bf R}]\right)^2}
{E_{S_1}[{\bf R}] - E_{S_0}[{\bf R}] + \alpha}.
\end{align}
The first term minimizes the average energy of the two states, while the second penalizes a nonzero energy gap $\Delta E = E_{S_1}-E_{S_0}$.
Here, $\sigma$ and $\alpha$ are hyperparameters. The parameter $\alpha$ regularizes the denominator of the penalty term, while $\sigma$ controls the strength of the constraint that drives the energy gap $\Delta E = E_{S_1}[{\bf R}] - E_{S_0}[{\bf R}]$ toward zero. 
In the limiting cases $\sigma \rightarrow \infty$ or $\alpha \rightarrow 0$, the minimum of the cost function converges to the true MECX energy. Importantly, the gradient of $F[{\bf R}]$ depends only on the individual state energy gradients and does not require derivative couplings, making this formulation directly compatible with our analytic gradient implementations.

In the present work, we adopt the recommended value $\alpha = 0.2$,
while $\sigma$ was adjusted manually during the optimization so as to enforce $\Delta E < 0.1$ mHartree at convergence.

As proof-of-concept benchmarks, we consider the twelve MECXs collected by Nikiforov \textit{et al.},\cite{Nikiforov14} spanning eight molecules. The set includes representative classes of CXs: twisted-pyramidalized intersections associated with partial homolytic bond breaking; twisted bond-length-alternating (tw-BLA) intersections characteristic of conjugated polyenes; and $n/\pi$ crossings such as those found in ethylene and methyliminium. Specifically, the MECXs investigated in this work comprise the twisted-pyramidalized (tw-pyr) MECX of ethylene ({\bf 1}), the ethylidene MECX of ethylene ({\bf 2}), the twisted bond-length-alternating (tw-BLA) MECX of methyliminium ({\bf 3}), the methylimine MECX of methyliminium ({\bf 4}), the MECX of ketene ({\bf 5}), the transoid MECX of butadiene ({\bf 6}), two tw-pyr MECXs of styrene ({\bf 7}, {\bf 8}),   the tw-BLA MECX of PSB3 ({\bf 9}), the tw-pyr MECX of stilbene ({\bf 10}), and two MECXs of the HBI anion ({\bf 11}, {\bf 12}).

For comparison, we use canonical methods suitable for multireference problems, namely MRCI and SF-TDDFT, whose results are reported in Ref.~[\onlinecite{Nikiforov14}], as well as SA-CASSCF results obtained in the present work. Among these, SA-CASSCF provides a useful mean-field reference, whereas MRCI and SF-TDDFT   incorporate dynamical correlation not explicitly taken into account in SACIS and SAECIS, but they should not be regarded as quantitatively converged reference standards.
In particular, collinear SF-TDDFT takes into account only the Hartree-Fock exchange contribution in the kernel, while neglecting semi-local exchange-correlation counterparts, leading to substantial functional dependence of the obtained results. Moreover, because the spin-flip excitation manifold does not span the full spin space, SF-TDDFT can suffer from spin contamination, which may affect the accuracy of CX energetics. For MRCI, the relatively small 6-31+G** basis set (6-31G** for stilbene and HBI due to the large computational cost) limits the completeness of dynamical correlation recovery. Furthermore, the calculations employ state-averaged CASSCF reference wave functions with molecule-dependent active spaces: (2e,2o) for methyliminium, stilbene, and anionic HBI, (4e,4o) for ethylene, butadiene, PSB3, and styrene, and (6e,5o) for ketene. The adequacy of these active spaces for fully describing the relevant near-degeneracies is not guaranteed.  
Nevertheless, SA-CASSCF provides a useful additional benchmark, while   MRCI and SF-TDDFT provide reasonable qualitative benchmarks against which the performance of the mean-field-based SACIS and SAECIS approaches can be assessed. It should also be noted that part of the observed structural differences relative to Ref.~[\onlinecite{Nikiforov14}] may arise not only from the underlying electronic-structure methods but also from differences in the optimization procedure, since the present SACIS and SAECIS calculations employ the Levine--Martinez penalty-function strategy, whereas Ref.~[\onlinecite{Nikiforov14}] used the branching-space update method\cite{Maeda10} for SF-TDDFT and analytic derivative couplings within a projected constrained optimization for MRCI.

Figure \ref{fig:CX} compares the optimized MECX geometries. Overall, both SACIS (blue) and SAECIS (red) are able to qualitatively reproduce the results of MRCI (gray) and SF-TDDFT (green). 
 For clarity, the SA-CASSCF structures are omitted from the superposition, because they are in most cases close to the MRCI geometries; the detailed comparison is discussed below.

\begin{figure*}
    \centering
\includegraphics[width=0.95
\linewidth]{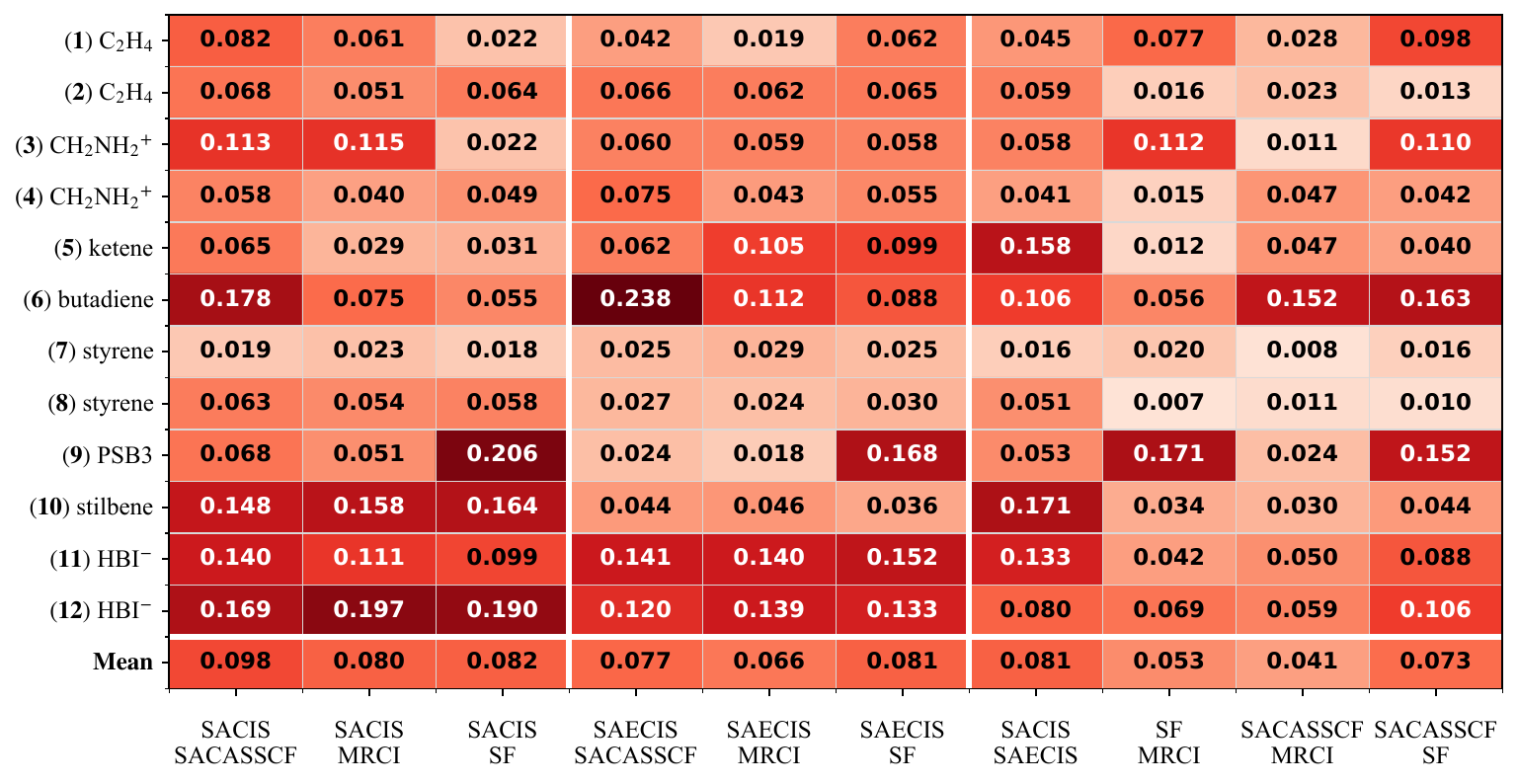}
    \caption{
    Pairwise RMSDs of the optimized MECX geometries obtained by SACIS, SAECIS, SA-CASSCF, MRCI, and SF-TDDFT. For the smaller systems ({\bf 1}--{\bf 6}), RMSDs were evaluated over all atoms including hydrogens, whereas for the larger conjugated systems ({\bf 7}--{\bf 12}) heavy-atom RMSDs were used to emphasize differences in the skeletal framework.}\label{fig:RMSD}
\end{figure*}
Figure~\ref{fig:RMSD} summarizes the RMSDs of the MECX geometries obtained with SACIS and SAECIS relative to SA-CASSCF,  MRCI, and SF-TDDFT. 
It also shows the differences between SACIS and SAECIS, between SF-TDDFT and MRCI, and between SA-CASSCF and the correlated references for consistency. To ensure a physically meaningful comparison, we adapt the RMSD calculation based on the system size and the nature of the reaction coordinates. For the smaller systems ({\bf 1}--{\bf 6}), all atoms including hydrogens are explicitly considered, as the displacements of hydrogen atoms---such as pyramidalization and twisting---are the primary drivers of the non-adiabatic coupling and characterize the electronic structure of these MECXs. In contrast, for the larger conjugated systems ({\bf 7}--{\bf 12}), the RMSDs are evaluated using heavy atoms to focus on the systematic deviations in the skeletal framework, where hydrogen motions are largely secondary to the deformation of the $\pi$-conjugated backbone.

Overall, the mean RMSDs indicate broadly comparable performance among the methods, consistent with Figure~\ref{fig:CX}; the average differences among the methods remain below 0.1~{\AA}. In many cases, such as {\bf 2}, {\bf 4}, {\bf 7}, and {\bf 8}, all methods reproduce very similar geometries. However, the table also indicates  that the degree of discrepancy is system-dependent. The additional SA-CASSCF comparison shows that SA-CASSCF reproduces the MRCI geometries reasonably well in most cases, except for butadiene, whereas its agreement with SF-TDDFT is generally somewhat poorer. For the tw-BLA methyliminium MECX ({\bf 3}), it appears the SACIS geometry is notably close to that of SF-TDDFT,  both of which deviate from MRCI by more than 0.1~{\AA}. SAECIS yields a geometry that is somewhat in the middle between MRCI and SF-TDDFT, as  suggested by  its nearly equal RMSDs relative to the two methods. 

In contrast, SA-CASSCF is much closer to MRCI (0.011 {\AA}) than to SF-TDDFT (0.110 {\AA}). More generally, SACIS does not simply follow the SA-CASSCF structures: the mean RMSD between SACIS and SA-CASSCF is comparable to those between SACIS and MRCI and between SACIS and SF-TDDFT. This suggests that, although SACIS is formally a mean-field method, its orbital-relaxed description does not merely reproduce the minimal-active-space SA-CASSCF picture, but can instead yield MECX geometries closer to either MRCI or SF-TDDFT depending on the system.

\begin{figure}
    \centering
    \includegraphics[width=1\linewidth]{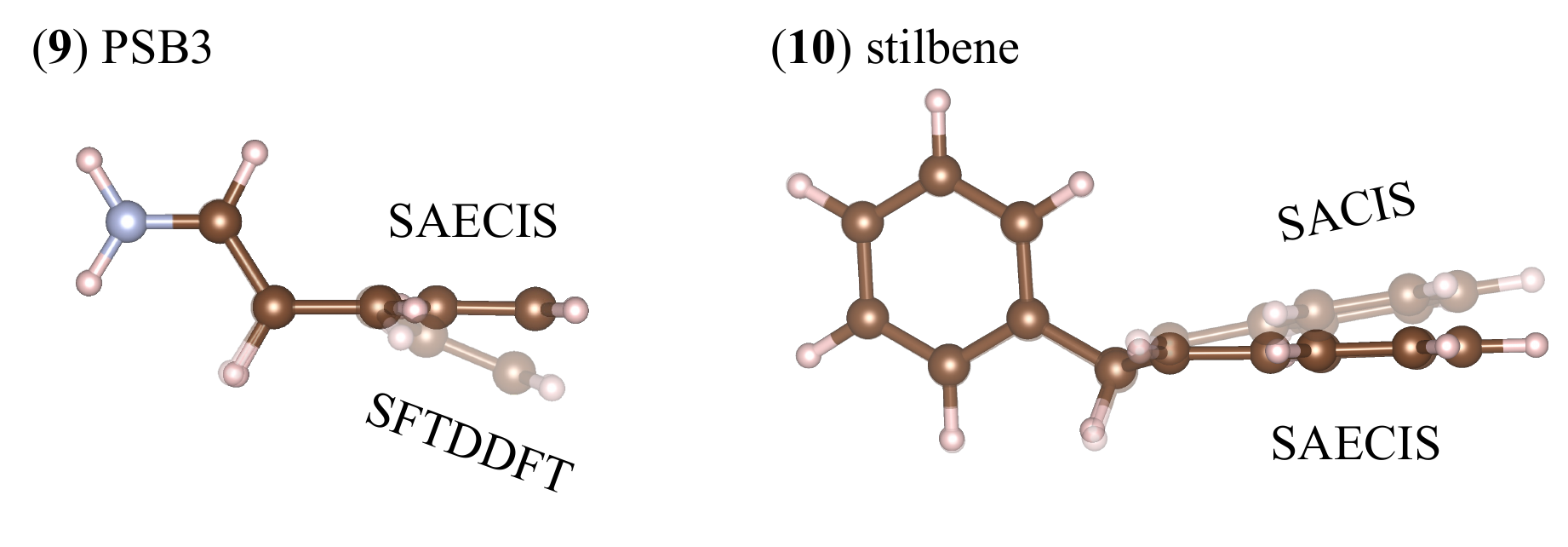}
    \caption{Comparison of MECX geometries for PSB3 (left) and stilbene (right). The opaque structures represent SAECIS results, while the transparent structures represent SF-TDDFT for PSB3 and SACIS for stilbene.}
    \label{fig:geom}
\end{figure}

For larger molecules, overall geometrical errors can often be traced to deviations in a few specific internal coordinates. In the case of PSB3 ({\bf 7}), SF-TDDFT exhibits relatively large deviations (0.168--0.206~{\AA}) compared to other methods, while the discrepancy among them remains sufficiently small (0.018--0.053~{\AA}). The dominant contributions arise from torsional mismatches in the central polyene segment as shown in Figure~\ref{fig:geom}: specifically, the torsion angles of the two central carbon-carbon bonds, $\phi$[(NH$_2$CH)CH=CH-CH=CH$_2$] and $\phi$[(NH$_2$)CH-CH=CH-CH(CH$_2$)], differ by -24.2$^\circ$ and +18.7$^\circ$, respectively. In contrast, the terminal iminium group torsion remains essentially unchanged ($\Delta \phi$[NH$_2$-CH=CH-CH(CHCH$_2$)] $ \approx 0^\circ$).
These results indicate that the RMSD is primarily driven by a redistribution of torsional distortion and conjugation within the central polyene, rather than by a rigid-body reorientation of the terminal iminium group.

Similarly, in the stilbene test ({\bf 10}), SACIS yields a slightly more twisted geometry than the spin-projected variant (Figure~\ref{fig:geom}). The structural difference is mainly characterized by the torsion angle of the Ph groups in Ph-CH=CH-Ph. The angle is 131.9$^\circ$ for SACIS, whereas 143.0$^\circ$, 144.5$^\circ$, and 141.1$^\circ$ for SAECIS, MRCI, and SF-TDDFT, respectively. This difference results in a slightly more tilted phenyl group in SACIS compared to other methods as shown in Figure~\ref{fig:geom}, yet yields marked deviations in terms of RMSD, 0.158--0.171~{\AA}.

It should be also mentioned that spin projection in SAECIS does not always guarantee the improved performance over SACIS for MECX search. Ketene and butadiene represent such examples. The overall accuracy in reproducing the geometries of MRCI and SF-TDDFT is similar between SACIS and SAECIS.

\begin{table*}[htbp]
\centering
\caption{Vertical excitation energies at the Franck--Condon (FC) geometry and adiabatic MECX energies (eV).
The FC values are defined as $\Delta E_{\mathrm{FC}} = E_{S_1}(\mathbf R_{S_0}^{\min})-E_{S_0}(\mathbf R_{S_0}^{\min})$.
The MECX values are adiabatic intersection energies defined as $\Delta E_{\mathrm{MECX}} = E_{\mathrm{CI}}(\mathbf R_{\mathrm{MECX}})-E_{S_0}(\mathbf R_{S_0}^{\min})$.}
\label{tb:energetics}
\begin{tabular}{llrrrrr}
\hline\hline
System & Geometry & SACIS & SAECIS & SA-CASSCF & MRCI & SF-TDDFT \\
\hline
C$_2$H$_4$ & FC & 7.54 & 7.96 & 6.87 & 8.11 & 7.70 \\
 & tw-pyr MECX ({\bf 1}) & 4.38 & 4.83 &  4.83 & 4.79 & 4.82 \\
 & ethylidene MECX ({\bf 2}) & 4.52 & 4.81 & 3.73& 4.69 & 4.52 \\
 
CH$_2$NH$_2^{+}$ & FC & 9.97 & 10.06 &  10.24 & 9.52 & 9.22 \\
 & tw-BLA MECX ({\bf 3}) & 4.57 & 4.15 &  4.11 & 3.82 & 4.27 \\
 & methylimine MECX ({\bf 4}) & 6.55 & 6.54 &  5.28& 5.63 & 5.46 \\

ketene & FC & 4.66 & 4.33 & 3.86&  4.01 & 4.01 \\
& MECX ({\bf 5}) & 3.41 & 2.14 & 2.54 & 2.43 & 2.82 \\

butadiene & FC & 6.76 & 7.40 &  6.20 &  6.89 & 5.94 \\
 & transoid MECX ({\bf 6}) & 5.31 & 5.56 &  5.63 & 4.89 & 5.19 \\

styrene & FC & 6.17 & 7.38 &  6.03& 5.76 & 5.29 \\
 & tw-pyr MECX ({\bf 7}) & 3.60 & 4.24 &  4.43&  4.68 & 4.01 \\
 & tw-pyr MECX ({\bf 8}) & 4.43 & 4.96 &  5.07&  5.28 & 4.83 \\

PSB3 & FC & 4.88 & 4.79 &  5.04& 4.47 & 4.43 \\
 & tw-BLA MECX ({\bf 9}) & 2.95 & 2.47 &  2.54 & 2.38 & 3.08 \\

stilbene & FC & 5.51 & 6.65 &  5.28 & 5.35 & 4.35 \\
 & tw-pyr MECX ({\bf 10}) & 3.58 & 4.45 &  4.50 & 4.50 & 4.20 \\

HBI$^{-}$ & FC & 4.11 & 2.61 &  4.43 & 3.88 & 3.14 \\
 & MECX ({\bf 11}) & 3.58 & 3.61 & 2.46& 2.87 & 2.84 \\
 & MECX ({\bf 12}) & 3.92 & 3.23 & 2.71& 3.20 & 3.20 \\
 \hline
 MAE & FC against MRCI &0.38 & 0.76 &  0.53 &  --- & 0.49 \\
     & MECX against MRCI &0.71& 0.34&   0.33& --- &0.31\\
     & FC against SF-TDDFT &0.73& 1.02 &  0.73 &0.49 &---\\
     & MECX against SF-TDDFT& 0.46 & 0.38 &  0.35 & 0.31 & ---\\
\hline\hline
\end{tabular}
\end{table*}
Finally, we investigate the relative energetics to assess the performance of the various methods.  Table~\ref{tb:energetics} summarizes the vertical excitation energies at the Franck--Condon (FC) geometry and the adiabatic MECX energies defined as the intersection energy relative to the ground-state minimum.

Because the results of MRCI and SF-TDDFT do not coincide quantitatively, any interpretation of the errors in CIS-based methods is necessarily reference-dependent. In particular, MRCI systematically predicts larger energy gaps than SF-TDDFT at both the FC and MECX geometries, with a non-negligible offset for several systems. Consequently, neither reference can be unambiguously regarded as a definitive energetic benchmark for the present data set. The discussion hereafter is therefore limited to a qualitative assessment of the energetic trends. Such an approach is appropriate, however, given that SACIS and SAECIS are inherently designed to provide a qualitatively correct description rather than high-precision quantitative accuracy. In this sense, the main value of SACIS and SAECIS in the present context lies in providing an efficient qualitative or semi-quantitative description of conical intersections, especially for structural optimization in larger systems, rather than in delivering quantitatively converged excitation energies.

This reference dependence is clearly reflected in the mean absolute errors (MAEs) of SACIS.
When measured against MRCI, the MAE at the FC point is relatively small (0.38~eV), whereas the MAE for the adiabatic MECX energies is substantially larger (0.71~eV).
However, when SF-TDDFT is taken as the reference, the apparent trend reverses: the FC MAE becomes larger (0.73~eV) than the MECX MAE (0.46~eV).
This inversion is not a physical inconsistency of SACIS per se, but rather a direct consequence of the systematic offset between the two correlated references.

By contrast, SAECIS and SA-CASSCF show a similar qualitative tendency: for both methods, the adiabatic MECX energies are reproduced more consistently than the FC vertical excitation energies. For SAECIS, the FC excitation energies exhibit several pronounced outliers, leading to relatively large MAEs despite the improved structural description. SA-CASSCF shows the same overall trend, with FC errors that remain moderate to substantial depending on the system, whereas the MECX energetics are in much better agreement with both correlated references on average. The fact that this behavior is common to both SAECIS and SA-CASSCF suggests that the improved energetic description at the MECX is primarily associated with balanced state-averaged orbital relaxation in the near-degenerate region, rather than with spin projection alone.

The observed disparity between the FC and MECX performance can be further understood by considering the role of orbital relaxation. As previously reported for valence excitations in systems with minimal static correlation,\cite{Tsuchimochi26} variational orbital optimization in CIS-based methods does not necessarily lead to a significant improvement in vertical excitation energies. At the FC geometry, the ground-state HF orbitals---optimized for the closed-shell $S_0$ state---provide a sufficient mean-field description, and the primary source of error remains the lack of dynamic correlation. However, the situation changes drastically when the excited-state nature deviates significantly from the ground-state HF mean-field, such as in Rydberg excitations or at heavily distorted geometries. In the case of Rydberg states, the diffuse nature of the electron density necessitates orbital readjustment that the ground-state orbitals cannot provide. Similarly, at the MECX, the molecular framework is significantly distorted and the $S_0$ and $S_1$ states become nearly degenerate. In this region, a balanced description of both states is paramount, requiring substantial orbital relaxation to account for the radical-like character and the change in bonding patterns. Our results suggest that  while SAECIS may not correct the vertical gaps at the FC point due to the inherent limitations of the CIS-based ansatz, its variational orbital optimization and spin projection become indispensable for capturing the physics of the MECX. The consistent energy gaps and structural agreement at these crossing points indicate that SAECIS successfully recovers the orbital relaxation effects necessary to describe the potential energy surface during large-amplitude nuclear displacements.

\section{Concluding remarks}
In this work, we have derived and implemented analytical nuclear gradients for SACIS and SAECIS and demonstrated their applicability to CX problems. The formulation required careful treatment of the singular electronic Hessian arising from redundant parametrization in the state-averaged CI space. By explicitly projecting out the null space of the CP equations, numerically stable and physically meaningful nuclear gradients were obtained, enabling reliable geometry optimization and MECX searches.

For the prototypical case of ethylene, both SACIS and SAECIS qualitatively reproduce the correct CX topology, in sharp contrast to conventional CIS and ECIS. The essential mechanism underlying this success is the fully variational orbital relaxation within the state-averaged framework, which alleviates ground-state orbital bias and effectively incorporates static correlation through symmetry breaking and localization effects in SACIS. Spin projection in SAECIS does not qualitatively modify the CX topology in this system, indicating that SACIS already provides an adequate zeroth-order description of the crossing seam.

The benchmark calculations on twelve MECXs further confirm that SACIS and SAECIS yield comparable structural accuracy, with mean RMSDs below 0.1~{\AA} relative to MRCI and SF-TDDFT. No clear and systematic quantitative performance difference between SACIS and SAECIS is observed for CX geometries or adiabatic MECX energetics. In this sense, both methods provide a qualitatively reliable description of CXs.

From a computational perspective, however, a distinction emerges. As demonstrated in the previous study, the inclusion of spin projection deteriorates the conditioning of the underlying equations, leading to slower convergence and increased numerical cost. Because the present results indicate essentially equivalent CX performance between SACIS and SAECIS, the additional overhead associated with spin projection does not translate into a systematic improvement in MECX accuracy. Therefore, in terms of cost-performance balance, SACIS is generally preferable: its qualitative performance matches that of SAECIS, while its computational efficiency is superior.

That said, SAECIS retains an important advantage in situations where higher-lying states with significant doubly excited character are involved, such as the $S_2$ state of ethylene and related systems. In such cases, the combination of orbital optimization and spin projection can effectively recover configurations inaccessible to pure single-excitation ans\"atze, providing a more balanced multistate description when three or more states are included in the state-averaging procedure.

Taken together, the present results establish SACIS as a computationally efficient and qualitatively reliable method for CX optimization, while positioning SAECIS as a valuable extension for cases requiring improved treatment of strong correlation and higher excited states.

Finally, the description of both SACIS and SAECIS remains to be qualitative, and quantitative accuracy is only achieved by inclusion of dynamical correlation. We are therefore currently working on such developments.

\section*{Acknowledgement}
This work was supported by JST FOREST Program, Grant No. JPMJFR223U, and JSPS KAKENHI, Grant No. 25K01733 and 25K22247.

\appendix
\section{Influence of the null space in the Z-vector to the nuclear gradients}
We consider how the redundant parametrization can affect analytic nuclear gradients for state-averaged CIS. 

Based on the discussion in the main text, the contribution of the null space ${\bf c}^J$ to the gradient is 
\begin{align}
    {\mathscr N}^{(x)} = \sum_{JK}\alpha_{JK} {\bf c}^{J,\dagger} \; {}^{\rm c}{\bf g}^{K,(x)} 
\end{align}
which can be shown to remain arbitrary.
To show the arbitrariness, we examine
\begin{align}
        {\bf c}^{J,\dagger} \;{}^{\rm c}{\bf g}^{K,(x)}
        &= \sum_\mu c^{J,*}_\mu \langle \Phi_\mu|(\hat H^{(x)} - E_{K}^{(x)})\hat P |0_K\rangle + h.c.
        \br&= \langle 0_J | (\hat H^{(x)} - E_{K}^{(x)}) \hat P |0_K\rangle + h.c.\br
        &= \langle 0_J|\hat H^{(x)} \hat P |0_K\rangle - E_K^{(x)} \delta_{JK} + h.c.
\end{align}
In general, $|0_K\rangle$, or its projected state, is not exact. Hence, the diagonal term
$\langle 0_K|\hat H^{(x)} \hat P |0_K\rangle - E_K^{(x)}$ does not vanish, i.e., the
Hellmann--Feynman identity does not apply. For $J\ne K$, the off-diagonal element
$\langle 0_J|\hat H^{(x)} \hat P |0_K\rangle$ is closely related to the nonadiabatic
(derivative) coupling: if $\hat P|0_K\rangle$ were an exact eigenstate of the Hamiltonian (which is not the case here),
\begin{align}
    \langle 0_J|\hat H^{(x)} \hat P |0_K\rangle
    = (E_K - E_J)\,\langle 0_J|\hat P |0^x_K\rangle \qquad (J\neq K).
\end{align}
In nondegenerate cases, if the Hellmann-Feynman identity holds for exact adiabatic eigenstates, the off-diagonal nonadiabatic coupling is generally nonzero, modulo symmetry-forbidden instances. Even when the identity breaks down for approximate states like CIS, introducing the electron-translation factor term cancels the non-Hellmann-Feynman contributions and restores the off-diagonal HF form, under which the coupling likewise remains nonzero unless prohibited by symmetry.

Therefore, ${\mathscr N}^{(x)}$ is generally non-zero unless $\alpha_{JK} = 0$; the null component has to be explicitly removed from the Z-vector to evaluate the nuclear gradients appropriately.

\bibliography{refs}
\end{document}